# Surface oxidation of liquid Sn

Alexei Grigoriev [a,*], Oleg Shpyrko [b], Christoph Steimer [a], Peter S. Pershan [a,b], Benjamin M. Ocko [c], Moshe Deutsch [d], Binhua Lin [e], Mati Meron [e], Timothy Graber [e], and Jeffrey Gebhardt [e]

[a] *Division of Engineering and Applied Sciences, Harvard University, Cambridge, Massachusetts 02138, USA*
[b] *Department of Physics, Harvard University, Cambridge, Massachusetts 02138, USA*
[c] *Department of Physics, Brookhaven National Lab, Upton, New-York 11973, USA*
[d] *Department of Physics, Bar-Ilan University, Ramat-Gan 52900, Israel*
[e] *The Consortium for Advanced Radiation Sources, The University of Chicago, Chicago, Illinois 60637, USA*

**Abstract**

We report the results of an x-ray scattering study that reveals oxidation kinetics and formation of a previously unreported crystalline phase of SnO at the liquid-vapour interface of Sn. Our experiments reveal that the pure liquid Sn surface does not react with molecular oxygen below an activation pressure of $\sim 5.0 \times 10^{-6}$ Torr. Above that pressure a rough solid Sn oxide grows over the liquid metal surface. Once the activation pressure has been exceeded the oxidation proceeds at pressures below the oxidation pressure threshold. The observed diffraction pattern associated with the surface oxidation does not match any of the known Sn oxide phases. The data have an explicit signature of the face-centred cubic structure, however it requires lattice parameters that are about 9% smaller than those reported for cubic structures of high-pressure phases of Sn oxides.

* Corresponding author. Present address: Harvard University, 9 Oxford Street, Cambridge, MA 02138, USA Phone: 1-617-495-4015, Fax: 1-617-496-4654
  *E-mail address:* alexey@xray.harvard.edu

# 1. Introduction

Chemical reactions at interfaces are of the fundamental and practical scientific interest. They sometimes exhibit both unusual kinetics and new phases that are unstable in the bulk [1-3]. There are considerable differences between the ways oxidation develops in various materials. One commonly cited example is the surface of solid aluminum. Although the oxidation of Al is rather rapid in the presence of even trace amounts of molecular oxygen, the formation of a relatively thin surface oxide layer effectively passivates the bulk from further oxidation [4]. By contrast, oxidation of metals like Fe proceeds well into the bulk. In spite of the fact that the free surfaces of liquid metals have recently attracted considerable attention because of the atomic ordering at the liquid-vapor interface [5-7] there have been very few studies of their reactive properties [8-11]. Oxidation of such surfaces are of particular interest because they lack the types of defects at which homogeneous nucleation occurs on solid surfaces namely steps, pits and dislocations [1]. In addition, surface oxidation of liquid metals can drastically change the surface tension which will have a profound effect on the way the liquid metal wets different surfaces [10]. This is important for practical processes such as soldering, brazing, casting etc.

The only two liquid metals for which the structure of the surface oxide has been studied by x-ray scattering technique are In and Ga, which were found to behave differently [8,9]. Oxidation of the liquid Ga surface is similar to that of solid Al in that it saturates at a 5 Å depth to form a uniform layer protecting the metal from further oxidation [12]. By contrast, oxidation of liquid In produces a rough oxide film from which there is negligible x-ray reflectivity signal [8]. Grazing incidence diffraction (GID) of the Ga surface oxide did not reveal any Bragg peaks, indicating that this oxide is amorphous. A direct comparison with In is not possible since GID measurements were not done for the surface oxide. In the present paper we report both x-ray reflectivity and GID studies of the oxide growth on the liquid Sn surface. In addition to the static features of the structure, these measurements also provide important information on the oxidation kinetics of the liquid Sn surface.

# 2. Background

## 2.1. Surface Scattering of X-Rays

X-ray reflectivity and GID are widely used for the determination of the structure of surfaces and interfaces, while off-specular diffuse scattering is used to probe surface homogeneity and roughness [13,14]. The present study makes use of all three techniques to characterize the oxidation of the liquid Sn surface. The geometry for these techniques is shown in the Fig. 1. X-rays of wavelength $\lambda$ and wave-vector $k = 2\pi/\lambda$ are incident at an angle $\alpha$ to the surface. For specular reflectivity the detected wave-vector is in the plane of incidence, $\Delta\Theta = 0$, at an angle $\beta = \alpha$ to the surface. The scattering is measured as a function of wave-vector transfer along the normal to the surface $q_z = 2k \sin\alpha$. For GID, the incident angle is generally less than the critical angle, $\alpha_c$, for total external reflection [14], and scattering is measured as a function of both the surface-parallel, $q_{xy} = k\sqrt{\cos^2\alpha + \cos^2\beta - 2\cos\alpha\cos\beta\cos(\Delta\Theta)}$ and surface normal, $q_z = k[\sin\alpha + \sin\beta]$, components of the wave vector. Small angle off-specular diffuse scattering is measured in the plane of incidence, $\Delta\Theta = 0$, as a function of $\beta$ for fixed $\alpha$ with $q_y = k[\cos\alpha - \cos\beta]$ and $q_z = k[\sin\alpha + \sin\beta]$.

The reflectivity $R(q_z)$ is commonly expressed as

$$\frac{R(q_z)}{R_f(q_z)} = |\Phi(q_z)|^2 \exp\left[-\sigma(q_z)^2 q_z^2\right],$$
$$\sigma(q_z)^2 = \sigma_{int.}^2 + \sigma_{cw}(q_z)^2, \quad (1)$$

where $R_f(q_z)$ is the Fresnel reflectivity that can be calculated from classical optics for a flat and structureless surface, $\Phi(q_z)$ is the surface structure factor, and $\sigma(q_z)$ is the effective surface roughness consisting of the intrinsic roughness $\sigma_{int}$ and the roughness that arises from thermally excited capillary waves $\sigma_{cw}(q_z)$ [8,13,14]. The $q_z$ dependence of $\sigma_{cw}(q_z)$ arises from the $q_z$ dependence of the reflectometer resolution [8]. If either the structure factor or the effective roughness is changed by the occurrence of a chemical reaction at the surface one should expect changes in the reflected signal.

There is a number of theoretical treatments of the small angle off-specular diffuse scattering for liquid surfaces [7,8,15]. From the general expression for the scattering cross section

$$\frac{d\sigma}{d\Omega} = \frac{A_0}{8\pi \sin\alpha} q_z^2 R_f(q_z) |\Phi(q_z)|^2 \frac{\eta}{q_{xy}^{2-\eta}} \left(\frac{1}{q_{max}}\right)^\eta \quad (2)$$

it can be seen that the scattering varies as $1/(q_{xy})^{2-\eta}$, where $\eta = [(k_B T)/(2\pi\gamma)] \cdot q_z^2$, $\gamma$ is the surface tension, and the upper cutoff of the capillary wave spectrum $q_{max} \approx \pi/a$, where $a$ is of the order of the atomic radius.

The theory for GID has been extensively reviewed [13,14,16]. For angles of incidence below the critical angle, x-rays penetrate evanescently from the surface into the bulk liquid and at sufficiently small angles the penetration depth is a few tens of angstroms. Thus, scattering observed as a function of $q_{xy}$ measures the surface-parallel structure within a thin surface layer only.

*2.2. Oxidation*

The formation of oxides on the liquid Sn surface can either occur through a homogeneous oxide nucleation at a clean part of the surface or inhomogeneously at nucleation centers preexisting at the sample's perimeter. For an oxide-free surface there is a possibility that some *threshold oxygen pressure* is necessary for homogeneous nucleation to occur in measurable times. For example STM studies of the initial stages of oxidation of the Si(111)-(7×7) reconstructed surface by Leibsle et al. [17] revealed that "… large defect-free areas on the surface are fairly insensitive to oxygen exposure…" with negligible oxidation after an oxygen exposure of 50 Langmuir (1 Langmuir = $10^{-6}$ Torr for 1 sec) [17]. They report that the STM

"images typically show no appreciable changes except for a few atomic sites or patches randomly distributed in an area of 500×500 Å$^2$." On the other hand they observe that "if the starting surface has a substantial number of defects … the overall oxidation rate is significantly enhanced." In view of the absence of defects on the free liquid metal surface a similar threshold effect on the surface of liquid Sn is plausible.

None of the previous oxidation studies of the liquid Ga or In surfaces have addressed the matter of an oxidation threshold. On exposure of a clean Ga surface to oxygen the first indication of oxides, as monitored by scanning focused ion-beam microscopy, appears to be the formation of small-size islands which subsequently grow to form irregular patches. These islands eventually cover the entire surface [18]. The remarkable feature of this oxide is the well-defined x-ray reflectivity signature indicating a uniform layer of ~5 Å thickness [9]. In contrast, the oxide forming on the surface of liquid In is sufficiently rough for x-ray reflectivity studies to be impractical [8].

## 3. Experimental details

The measurements presented here were carried out at ChemMatCARS 15-ID-C experimental station of the Advanced Photon Source at Argonne National Lab at an x-ray wave-length of 1.1273Å (11.0 KeV). A Sn sample of 99.9999% purity was placed in a molybdenum pan (diameter ~60 mm) and heated under UHV conditions to $240^0$ $C$ just above the Sn melting point $T_m = 232^0$ $C$. The vacuum in the baked out chamber was in the $10^{-10}$ Torr range and the oxygen partial pressure was below $10^{-11}$ Torr. Visible patches of native oxide present in the original sample were removed by a combination of mechanical scraping and sputtering with Ar$^+$ ions. The cleaning process was monitored visually with CCD camera and by x-ray reflectivity measurements. X-ray reflectivity from the Sn surface initially covered with rough native oxide is too weak to measure above the critical angle. An oxide-free region of at least ~50 mm diameter was obtained after sputtering for approximately 48 hours. The radius of the surface curvature at the center of the cleaned sample was determined to be ~15 m by measurements of the reflection angle $\beta$ for a fixed nominal incidence angle $\alpha$ at different points on the surface [19].

In a separate publication [5] we report x-ray studies of the oxide-free liquid Sn surface that was achieved following this cleaning process. The reflectivity, $R(q_z)$ was measured before the oxidation to a maximum of $q_z$ ~2.6 Å$^{-1}$ [5]. The surface layering that was observed is similar to that previously reported for the surfaces of liquid Ga, In, Hg and K [6,7,19]. A surface layering anomaly observed for liquid Sn, as well as measurements carried out to establish the atomic purity of the liquid Sn surface are discussed in that publication [5]. The study of atomically clean liquid Sn surface was followed by controlled oxidation of the cleaned Sn sample through introduction of ultra-high purity research grade oxygen gas (99.9999% purity, Matheson Tri-Gas Inc.) through a leak valve. The oxidation was done in several steps by exposing the UHV chamber to selected oxygen pressures for fixed time intervals. After obtaining the desired dosages the oxygen was pumped out of the chamber and measurements were carried out in an oxygen-free environment.

## 4. Results and discussion

*4.1. Oxidation kinetics*

The steps of oxygen exposure are listed in Table 1. The complete specular x-ray reflectivity scan was taken prior to oxidation, then measured again after the sum of exposures in steps #1 and #2 described in Table 1, and finally after the complete oxidation of the sample, when the surface visually appeared to be heavily oxidized. These results are shown in Fig. 2. The inset of Fig. 2 displays the same data normalized by the Fresnel reflectivity, $R_f(q_z)$, for pure Sn. In addition to these reflectivity scans the reflected signal at the point corresponding to the specular condition at $q_z$ = 0.5 Å$^{-1}$ was continuously monitored during the entire oxidation process even when the Sn surface was heavily oxidized and the specular signal was smeared out.

Fig. 2 depicts the reflectivity from the clean and oxidized surface of liquid Sn. The inset shows that for $q_z$ < 1 Å$^{-1}$ the specular x-ray reflectivity recorded from the oxidized sample at step #3 has essentially the same shape as that of the unoxidized clean Sn surface. The region of sample's surface that is probed by these two measurements is the area illuminated by the incident beam on the sample surface, so-called *"footprint"*. Its length is equal to (*Beam Height*)/sin($\alpha$). This value decreases with increasing $q_z$ from 2.2 mm for $q_z \approx$ 0.1 Å$^{-1}$ to 0.22 mm when $q_z \approx$ 1 Å$^{-1}$. Reflectivities measured at several different positions on the sample surface were found to be virtually identical with that shown in Fig. 2. Therefore, we conclude that the oxidation effect at step #2 is to *homogeneously* reduce the reflectivity for $q_z$ < 1 Å$^{-1}$ by a factor of ~0.45. The simplest explanation that would account for this is that the oxidation proceeds by formation of oxide patches which coexist with clean Sn surface patches in a way similar to that observed for liquid Ga [20]. With this interpretation at step #3 about 55% of the surface would be coated by a non-reflecting oxide layer, while the remaining 45% would be clean liquid Sn patches.

Small angle off-specular diffuse scattering data are presented in Fig. 3 for the clean Sn surface and for the partially oxidized surface at step #3. The measured points for the clean liquid surface (open circles) agree well with the $1/(q_{xy})^{2-\eta}$ dependence (line) predicted by Eq. 2 for a liquid surface [5]. The shape of the scan for the partially oxidized surface differs from that of the clean surface only over the small region in the vicinity of the specular peak, $q_{xy} \approx$ 0 (i.e. $\beta \approx \alpha$). This requires a slight revision of the interpretation of the oxidation effect, since if 55% of the surface had been covered by non-reflecting patches, as suggested above, the $|q_{xy}| > 0$ wings of the diffuse scattering curve would have to be also lower than those of the clean surface. The present observation that the intensity of the diffuse wings for the oxidized sample is the same as that of the clean surface can be explained by assuming that the oxide patches cover a much smaller fraction of the surface than the 55% mentioned above, but they distort the liquid surface locally. For the diffuse scattering the detector slit is only 1 mm high, corresponding to angular resolution of 0.09 degrees, while for the specular reflectivity it was 4 times larger. Local deviations of the surface from the horizontal by 0.2 degrees for ~50% of the surface would be sufficient to reduce the measured intensity of the sharp specular peak without having a measurable effect on the broadly-distributed off-specular diffuse scattering. By this interpretation the fraction of the surface that is actually oxidized could be less than ~55%.

Fig. 2 also shows that the reflected signal from the completely oxidized sample (step #8) decreases rapidly with increasing $q_z$ and becomes unmeasurably small at $q_z \cong 0.3$ Å$^{-1}$. Between steps #5 and #8 the sharp specular peak at $\beta = \alpha$, $\Delta\Theta = 0$ has disappeared. This occurs because the oxidation induces a surface roughness that smears out the specular peak beyond the resolution of the reflectometer. This evolution of the surface oxide is consistent with the previous results for oxidized surface of liquid In [8]. In fact this behavior indicates that the mean square roughness of the oxide grown at the Sn surface at this step is greater than 5 Å. In the following, we will continue to refer to the signal as reflectivity despite the loss of a distinct specular peak.

Regardless of whether or not exactly 55% of the surface is oxidized after the total exposure of $3.1 \times 10^3$ L at $2 \times 10^{-5}$ Torr following the steps #1 and #2, it is reasonable to suggest that the specular reflectivity at some fixed $q_z$ is proportional to the fraction of the surface that is coated with oxide. The data in Fig. 4 show the signal at $q_z = 0.5$ Å$^{-1}$ as a function of time corresponding to the entries in the Table 1. The inset of Fig. 4 shows the slopes $\frac{1}{R/R_f} \frac{d}{dt}\left[\ln(R/R_f)\right]$, which we designate as our "oxidation rate", as a function of the oxygen pressure. This "oxidation rate" clearly shows a monotonic dependence on the oxygen pressure, as well as a clear oxidation threshold pressure of $\sim 5 \times 10^{-6}$ Torr.

Before the threshold oxygen pressure was applied there was no variation in the reflectivity at $q_z = 0.5$ Å$^{-1}$ during the oxygen exposure at $5 \times 10^{-6}$ Torr (step #1.3) for 46 minutes. It is conceivable that during this oxygen exposure the surface area of the sample ($0.4 \times 4$ mm$^2$) probed by x-rays was swept clean of homogeneously nucleated oxides, for example, by the Marangoni effect. On the other hand there are good reasons to rule out the effects "hiding" an oxide growth below the threshold oxygen pressure. The first is that in addition to the scans shown in Fig. 2 a number of specular reflectivity scans were measured in the vicinity of the critical angle at oxygen pressures $<2.0 \times 10^{-5}$ Torr. At this incidence angle the sample surface area illuminated by the incident x-ray beam is of the order of $5 \times 4$ mm$^2$ that is about 12 times larger than the area probed at $q_z = 0.5$ Å$^{-1}$. None of the reflectivities showed any indication of transient signals that could be interpreted as coming from oxide patches flowing through the illuminated region. The second reason is that upon raising the oxygen pressure to $2.0 \times 10^{-5}$ Torr the reflected signal at $q_z = 0.5$ Å$^{-1}$ decreased by a factor of 2 in about 2 minutes (compare with 46 minutes at $5 \times 10^{-6}$ Torr). As can be seen from Fig. 2 this reduction is homogeneous across the illuminated area and it doesn't seem likely that this sudden reduction was caused by an inhomogeneous oxidation front sweeping across the illuminated area.

After oxidizing the sample at step #2 the oxygen was pumped out of the chamber and the reflectivity scan was taken at pressure in $10^{-10}$ Torr range. Then the oxygen pressure was raised again to $1.0 \times 10^{-6}$ Torr (step #4) that is below the threshold. As shown in the Fig. 4, the rate at which the reflected intensity falls in step #4 is roughly an order of magnitude less than the rate at $2.0 \times 10^{-5}$ Torr. At this point the surface is partially oxidized and the observed rate for further oxidation is likely to arise from some average between oxidation that is induced by the "old" oxide residing at the perimeter and "new" oxide due to homogeneous nucleation at the clean surface. In view of the fact that the rate for homogeneous nucleation of the clean portion of the surface is certainly going to be a small fraction of the rate associated with oxidation due to existing surface oxides, it is hard to avoid the conclusion that there is a sharp non-linear change in the rate of homogeneous oxidation between $1.0\times10^{-6}$ Torr and $2.0\times10^{-5}$ Torr (see inset of Fig. 4). In fact, assuming that the rate of change of $\frac{1}{R/R_f}\frac{d}{dt}\left[\ln(R/R_f)\right]$ is a monotonic function of the oxidation rate, the data shown in the Fig. 4 implies that the non-linear change in oxidation rate occurs between step #5, where the pressure is $3.0\times10^{-6}$ Torr, and step #6 where the pressure is $6.2\times10^{-6}$ Torr. Thus, it seems most likely that the threshold for homogeneous oxidation falls within this pressure range.

*4.2. Oxide Structure*

The structure of the Sn surface oxide was probed by GID measurements. The results for the liquid Sn surface before oxidation and after an exposure of 4000 L (during step #3) are presented in Fig. 5 together with a plot of the difference between the two. In addition, simulated diffraction patterns of known Sn oxide structures are presented in Fig. 5 below the experimental data. The broad peak around $q_{xy}$ = 2.234 Å$^{-1}$ corresponds to the bulk liquid structure of Sn and is in good agreement with published data [21]. The two sharp oxide peaks appear at $q_{xy}$ = 2.442 Å$^{-1}$ and 2.827 Å$^{-1}$. No other peaks were observed in the accessible range $0 \leq q_{xy} \leq 3.2$ Å$^{-1}$. The positions of the observed peaks change with $q_z$ (in the experiment $\beta$ was varied at a fixed $\alpha$) and follow the powder diffraction pattern cone as shown in Fig. 6. This result indicates that the new peaks appeared as a result of the Sn surface oxidation formation of a three dimensional powder crystal structure.

The possibility that these GID line originate in a contaminant, e.g. an oxide of a different metal or a salt of Sn other than an oxide, is highly unlikely considering the null result of an exhaustive search for such contaminants carried out for the unoxidized liquid Sn surface [5], as mentioned above, and the high purity of the oxygen used. Thus, one has to conclude that the GID lines belong to a Sn oxide grown at the liquid Sn surface, and that this oxide has a structure which is different from that of the known bulk Sn oxides at ambient conditions.

The two most likely surface oxides expected for Sn are SnO and $SnO_2$, which are the only bulk oxides of Sn observed at ambient conditions. At ambient pressure and temperature both of these have a tetragonal structure but the dioxide has a more isotropic structure with alternating atomic planes of Sn and O. The monoxide has a layered structure with the atomic planes of the same species adjacent each other. The crystalline structures of SnO and $SnO_2$ belong to space groups *P4/nmm* (romarchite) and *P4$_2$/mnm* (rutile), respectively [22]. The expected Bragg peak positions in Fig. 5 clearly do not correlate with the crystalline structure of the surface oxide of liquid Sn. Moreover, it is not possible to match these structures to that of Sn surface oxide by lattice parameter adjustment within a reasonable range (variations of up to 20% from nominal values).

There are reported high-pressure modifications of the Sn oxides' structure [23-25]. Two of them are cubic structures belonging to $Pa\overline{3}$ (primitive cubic) and $Fm\overline{3}m$ (fluorite type, rock salt) space groups. The transition to these structures occurs at pressures above 21 GPa. The diffraction patterns of these structures are presented in Fig. 5. These patterns also do not match the peaks experimentally observed for the surface Sn oxide; however, the most significant observation is that the ratio of positions of the two diffraction peaks of the surface oxide is $\sqrt{3}/2$, which is precisely the ratio of the two lowest order peaks for the high-pressure phases. In fact, to the best of our knowledge the face-centered cubic (fcc) lattice is the only one for which the two smallest reciprocal lattice distances are in this ratio. Thus it is not likely that the observed surface oxide is anything other than the fcc structure. The problem with this is that in order to produce the observed surface oxide peaks the lattice parameter must have a value of 4.455 Å which represents a 9% reduction from the published values of 4.87 Å and 4.925 Å for the high-pressure structures $Pa\bar{3}$ and $Fm\bar{3}m$ respectively. Assuming that all of the appropriate fcc lattice positions are occupied by Sn atoms, the calculated density of the lattice is between 1.5 and 1.8 times higher than the densities of the known oxides [25]. At first glance it seems rather implausible; however, the Sn-O closest neighbor distance of this form of the surface oxide is 1.83 Å for the $Pa\bar{3}$ structure and 2.23 Å for the rock salt structure. The latter precisely matches the molecular Sn-O bond length of the romarchite SnO structure (2.20 Å). On the other hand, since the former is smaller than the molecular Sn-O covalent bond length of 1.95 Å, the rock salt structure may be more realistic. Moreover, the rock salt structure along the [111] crystallographic direction consists of alternating planes of Sn and O atoms with a spacing between two closest Sn planes of 2.57 Å which matches the spacing found between uppermost surface layers of the clean liquid Sn surface of 2.55 Å [5]. A second argument in favor of the rock salt structure is that the (200) to (111) GID peak intensity ratio of the rock salt structure, 0.7, is significantly closer to the experimentally observed ratio, 0.8, than the ratio of peaks of the $Pa\bar{3}$ structure 0.57. Thus, although the density of the fcc form of the Sn-oxide is high, the lattice spacings themselves are not unphysical.

Without measurements of higher order Bragg peaks our interpretation of a high density fcc surface oxide can be considered somewhat speculative. On the other hand, in view of the fact that the only two small angle Bragg peak positions appear in exactly the ratio for the fcc lattices it is very difficult to think of an alternative lattice.

## 5. Summary

In the present study we found that the surface of liquid Sn is resistant to oxidation until a threshold oxidation pressure is reached. The oxidation begins within the pressure interval of $3.0 \times 10^{-6}$ Torr and $6.2 \times 10^{-6}$ Torr. Above this threshold oxidation pressure a uniform growth of oxide islands on the liquid Sn surface is detected. The islands' surface is very rough and the presence of oxide patches is manifested in a uniform reduction of the specular signal reflected by clean Sn surface.

The diffraction pattern associated with the surface oxide does not match any of the known Sn oxide phases. On the other hand, the appearance of only two Bragg peaks whose scattering vectors are in the ratio of $\sqrt{3}/2$, and with the observed intensities is an unambiguous signature of the fcc structure. It is troublesome that the calculated density for this fcc phase is 50% to 80% higher than that of the published Sn oxide phases; however, we have not been able to construct an alternative identification. It would be very useful to extend the GID measurement to search for higher wave-vector Bragg peaks.

These observations raise the question of whether the structures that form in a chemical reaction at the surfaces of a liquid metal are fundamentally different from those that form in a bulk chemical reaction, or at the surface of a crystal. For example, we do yet not know the structure of oxides that form on the various facets of a Sn crystal. Are any of them similar to the surface oxide that forms on liquid Sn? Secondly, is the anomalous surface oxide on liquid Sn an indicator that anomalous oxides will also form on the surfaces of other liquid metals? We know that the surface oxide on liquid Ga is amorphous but we do not know if the rough oxide that forms on the surface of liquid In is crystalline or not. This is clearly an area that requires further study. From a theoretical perspective there is a compelling need to ask why this particular cubic phase is favored by the surface over the "normal" structures of the oxides.

**Acknowledgements**

This work has been supported by the U.S. Department of Energy Grants No. DE-FG02-88-ER45379 and DE-AC02-98CH10886, the National Science Foundation Grant No. DMR-01-12494, and the U.S.-Israel Binational Science Foundation, Jerusalem. ChemMatCARS Sector 15 at the Advanced Photon Source is principally supported by the National Science Foundation/Department of Energy under the grant number CHE0087817. The Advanced Photon Source is supported by the U.S. Department of Energy, Basic Energy Sciences, Office of Science, under Contract No. W-31-109-Eng-38.

**Figure and table captions.**

Fig. 1. Kinematics of the x-ray scattering used in the present study.

Fig. 2. X-ray reflectivity of the clean liquid Sn surface (•) (step #0); the surface after an oxygen dose of $3.1\times10^3$ L at an oxygen partial pressure $2\times10^{-5}$ Torr ($\nabla$) (step #3) and after the complete oxidation of the surface ($\square$) (step #8). The inset shows the measured reflectivities normalized by the Fresnel reflectivity.

Fig. 3. Off-specular diffuse scattering scans from the clean liquid Sn surface (circles) (step #0) and after an exposure to the oxygen dose of $3.1\times10^3$ L at an oxygen partial pressure $2\times10^{-5}$ Torr (triangles) (step #3). Scans are recorded at $q_z = 1.0$ Å$^{-1}$.

Fig. 4. Fresnel-normalized x-ray reflected intensity at $q_z = 0.5$ Å$^{-1}$ during the oxidation process as a function of time. The inset shows the relative rate of change of the reflected intensity, $\frac{1}{R/R_f}\frac{d}{dt}[\ln(R/R_f)]$, which qualitatively reflects the oxidation rate. Circles in the inset denote stable logarithmic rates and triangles represent the rates which are changing with time. Large filled circles in the figure (#0, #3, #8) denote points where complete reflectivity measurements were taken. Other regions are: #1 — oxygen pressures below $5.0\times10^{-6}$ Torr; #2 — $2.0\times10^{-5}$ Torr; #4 — $1.0\times10^{-6}$ Torr; #5 — $3.0\times10^{-6}$ Torr; #6 — $6.2\times10^{-6}$ Torr; #7 — $1.0\times10^{-5}$ Torr; thick solid lines denote regions where oxygen was not applied, but measurements were continued.

Fig. 5. Grazing Incidence Diffraction (GID) scans of the clean liquid Sn surface and the surface after exposure to a total oxygen dose of $3.1\times10^3$ L as described in step #3 of Table 1. The difference plot highlights the surface oxide diffraction pattern showing two peaks at $q_{xy} = 2.442$ Å$^{-1}$ and 2.827 Å$^{-1}$. Positions of the diffraction lines of known SnO and SnO$_2$ structures are shown in the figure along with their crystal models.

Fig. 6. GID peak position dependence as a function of $q_z$. Simulation of the powder diffraction pattern cone is shown as a solid line.

Table 1. The sequence of oxidation and measurements on the liquid Sn surface.

**Figures and table.**

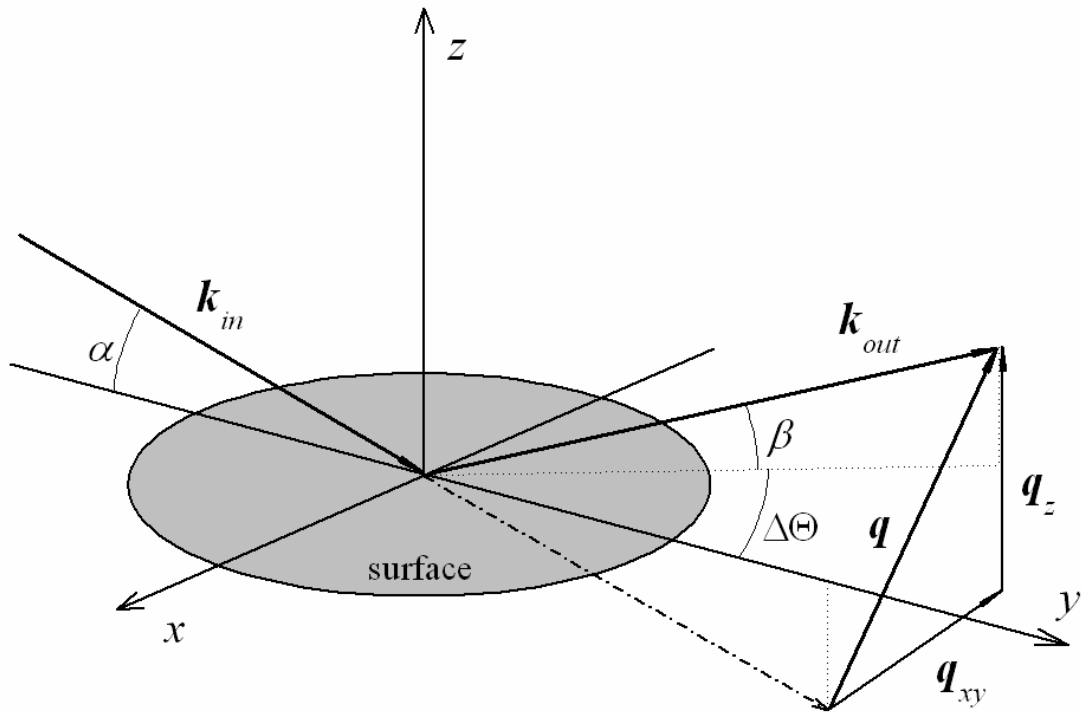

Fig. 1. Grigoriev et al.

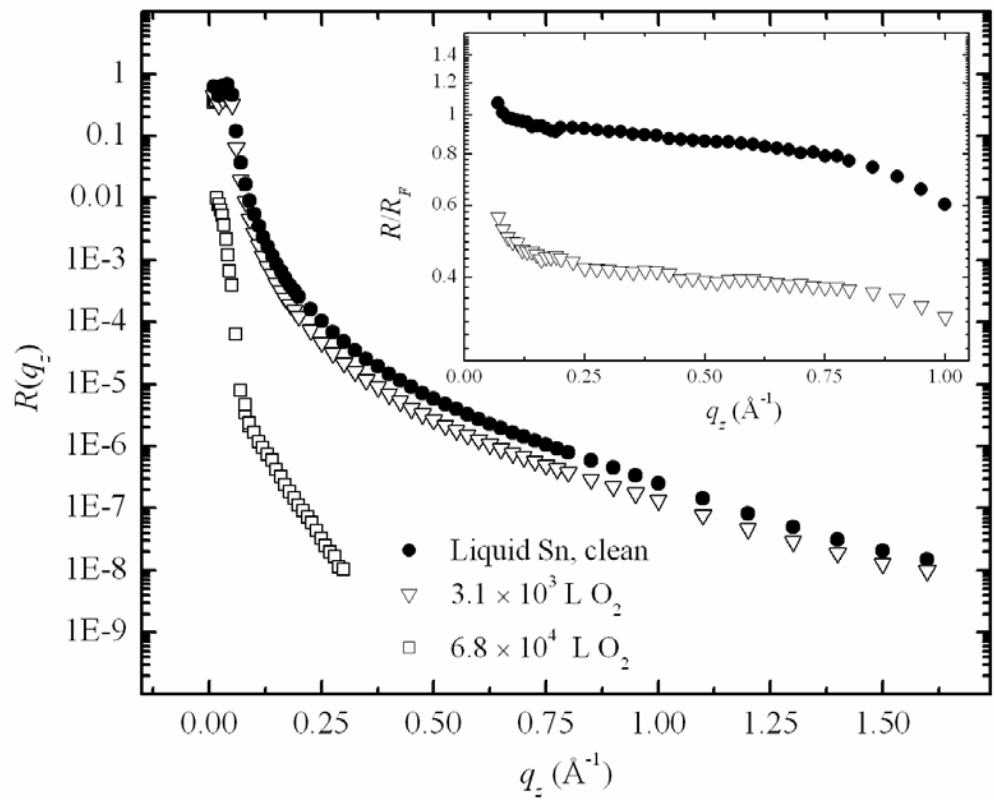

Fig. 2. Grigoriev et al.

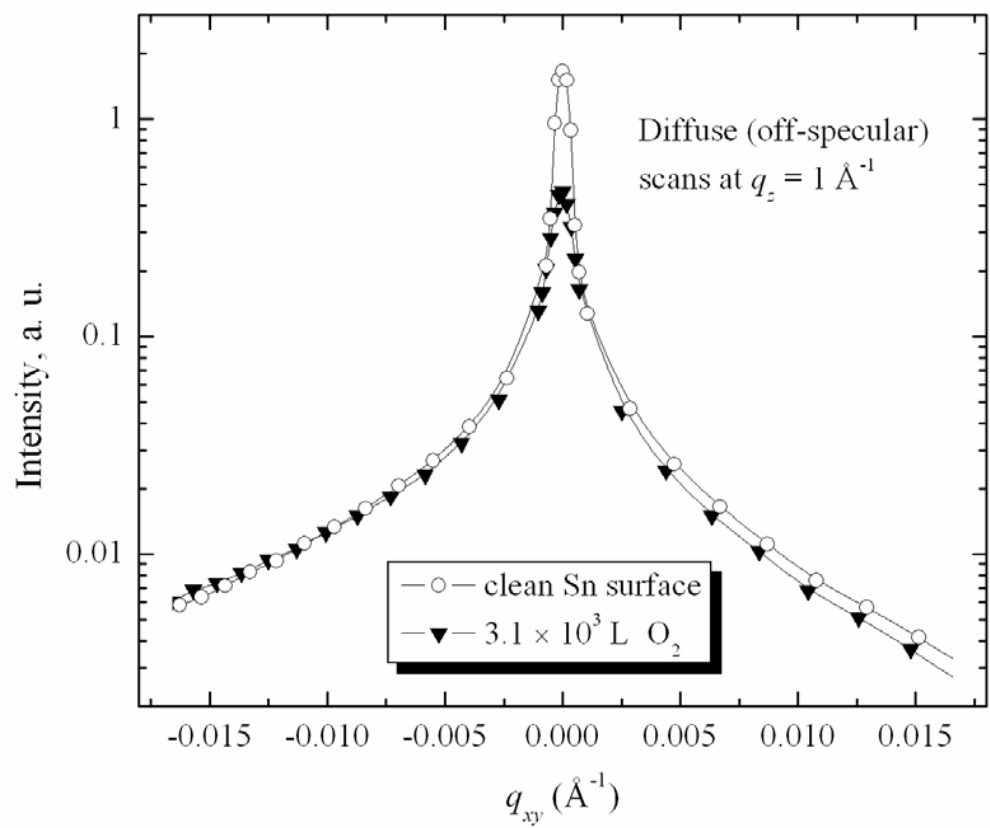

Fig. 3. Grigoriev et al.

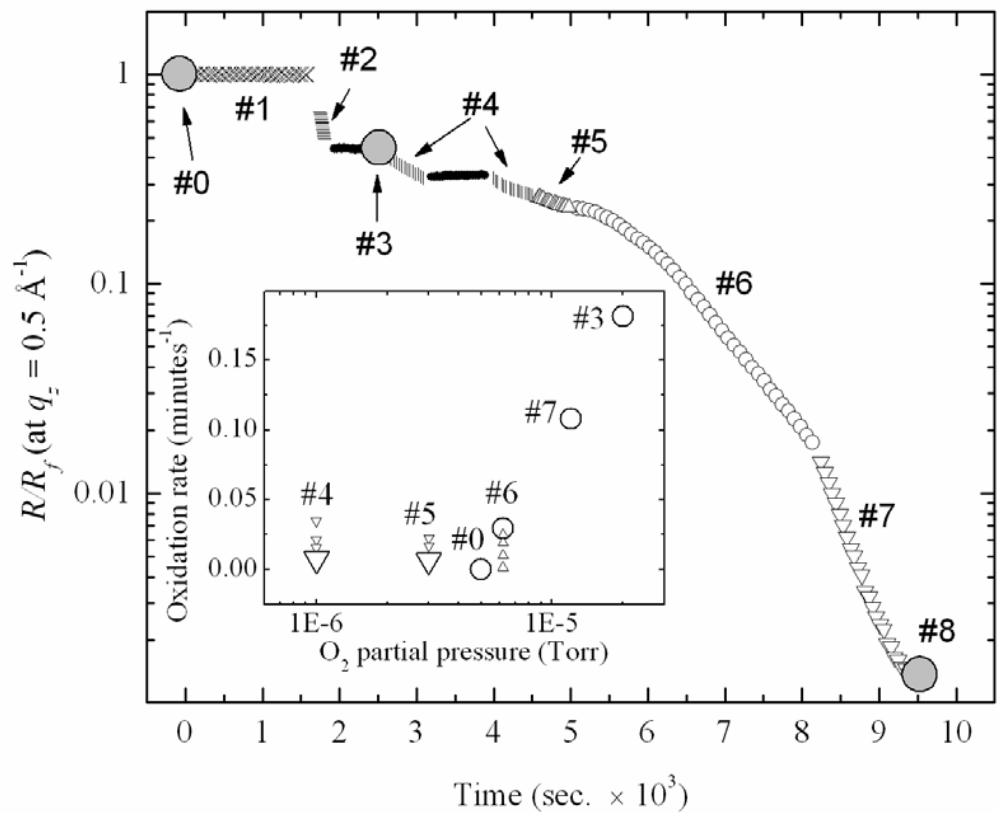

Fig. 4. Grigoriev et al.

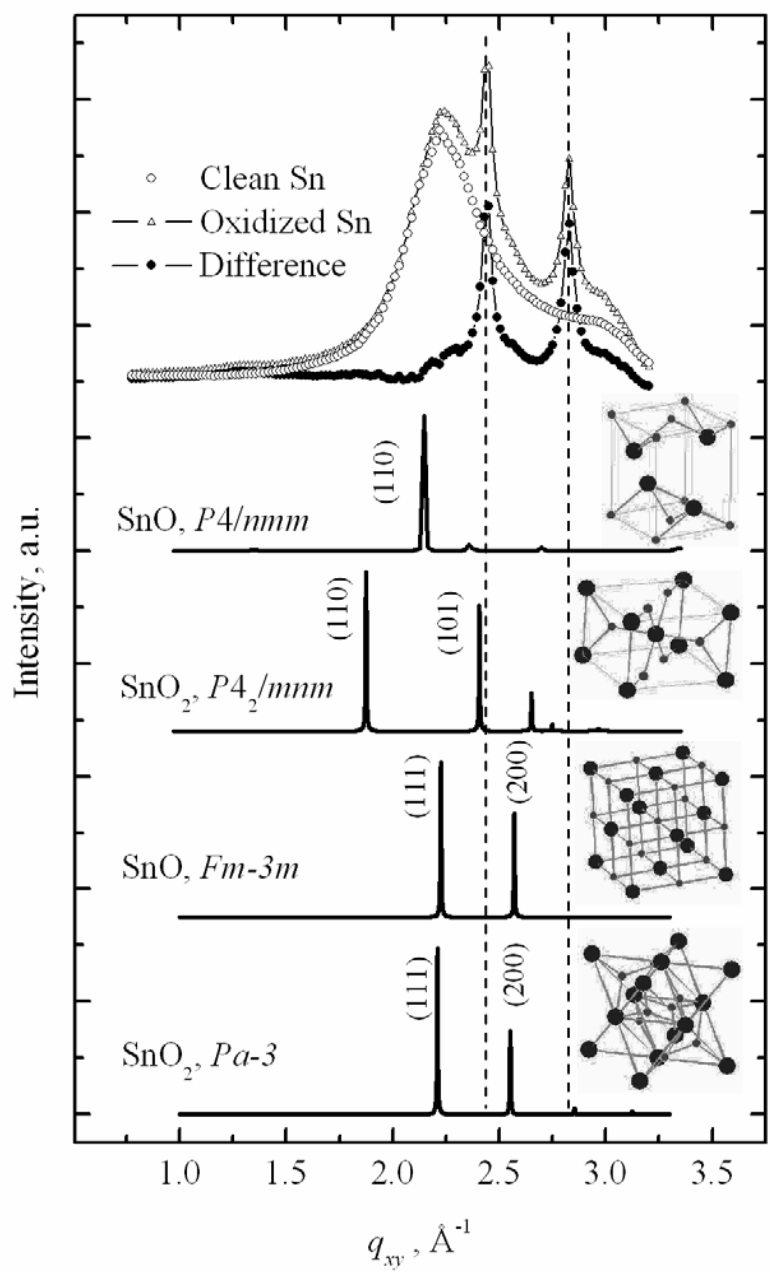

Fig. 5. Grigoriev et al.

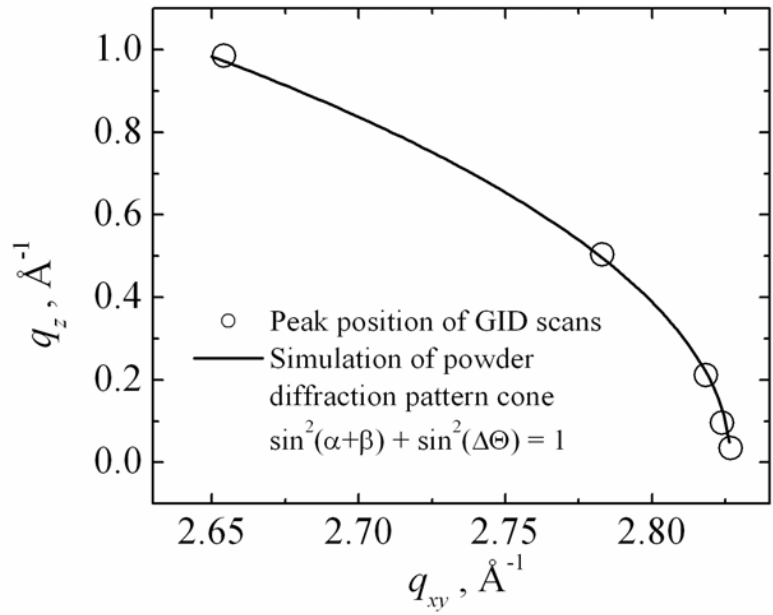

Fig. 6. Grigoriev et al.

Table 1. Grigoriev et al.

| Step Number | $O_2$ Pressure (Torr) | Time of Exposure (minutes) | Exposure/Total Exposure (Langmuirs) |
|---|---|---|---|
| #0 Complete Reflectivity | UHV, no oxygen | > 24 hours | — |
| #1 Dynamic measurement at $q_z=0.5$ Å$^{-1}$ | (#1.1) $5\times10^{-7}$<br>(#1.2) $1\times10^{-6}$<br>(#1.3) $5\times10^{-6}$ | 2<br>16.5<br>46 | 60 / 60<br>990 / 1050<br>13800 / 14850 |
| #2 Dynamic measurement at $q_z=0.5$Å$^{-1}$ | 2.0 x $10^{-5}$ | 2.6 | 3120 / 17970 |
| #3 Complete Reflectivity | UHV, no oxygen | — | — |
| #4 Dynamic measurement at $q_z=0.5$ Å$^{-1}$ | (#4.1) $1\times10^{-6}$<br>(#4.2) $1\times10^{-6}$ | 10.6<br>9.5 | 636 / 18606<br>570 / 19176 |
| #5 Dynamic measurement at $q_z=0.5$ Å$^{-1}$ | $3\times10^{-6}$ | 8 | 1440 / 20616 |
| #6 Dynamic measurement at $q_z=0.5$ Å$^{-1}$ | $6.2\times10^{-6}$ | 52.5 | 19530 / 40146 |
| #7 Dynamic measurement at $q_z=0.5$ Å$^{-1}$ | $1.2\times10^{-5}$ | 38.6 | 27792 / 67938 |
| #8 Complete Reflectivity | UHV, no oxygen | — | — |